# Domain decomposition and locality optimization for large-scale lattice Boltzmann simulations

M. Wittmann[a], T. Zeiser[a], G. Hager[a], G. Wellein[a,b]

[a]*Erlangen Regional Computing Center, University of Erlangen-Nuremberg, Germany*
[b]*High Performance Computing, Department of Computer Science, University of Erlangen-Nuremberg, Germany*

**Abstract**

We present a simple, parallel and distributed algorithm for setting up and partitioning a sparse representation of a regular discretized simulation domain. This method is scalable for a large number of processes even for complex geometries and ensures load balance between the domains, reasonable communication interfaces, and good data locality within the domain. Applying this scheme to a list-based lattice Boltzmann flow solver can achieve similar or even higher flow solver performance than widely used standard graph partition based tools such as METIS and PT-SCOTCH.

## 1. Introduction

The lattice Boltzmann method (LBM) [11] is used in computational fluid mechanics (CFD) to simulate incompressible athermal flows. In this work, the two-relaxation-time (TRT) collision model [5] is applied on a D3Q19 lattice. This means simulation takes place in a 3-dimensional space and each node holds 19 particle distribution functions (PDFs) and points to 18 surrounding neighbors.

Domain decomposition of an equidistant Cartesian mesh seems to be a trivial task. However, partitioning complex geometries (e.g. porous media) on large processor counts is a challenge when the resulting partitions should contain an equal amount of work and a low number of interface cells to neighbor partitions. The lack of scalable tools, their high memory consumption and long run-times of the partitioner make domain decomposition a tedious task. Moreover, most tools only return a cell-to-partition mapping but not the order in which the cells within a partition should be processed later on in the flow solver.

In this paper we present an approach to address these challenges. We start in Sec. 2 with a short overview of different methods for representing the data structures of the fluid in the flow solver. Here we concentrate on the so called *sparse representation*. Next in Sec. 3, we describe the steps needed to setup this representation within our "preprocessor" as well as with METIS and PT-SCOTCH (Sec. 4). Finally, in Sec. 6 and 7, performance results for the preprocessor and the flow solver are summarized that were obtained from our test bed (Sec. 5) and an outlook to future work is given in Sec. 8.

## 2. Data structures of the LB flow solver

Several options exist for representing the data structures of the fluid inside the flow solver. Note that this representation may be different from the one used for partitioning/preprocessing step.

Simple lattice Boltzmann implementations use the marker-and-cell (MAC) method on equidistant Cartesian meshes. Both the flag field used to distinguish fluid from obstacle cells and the particle distribution functions (PDFs) of the lattice Boltzmann method, are stored in multi-dimensional arrays [14]. We call this a *full arrays* approach since solid parts are just skipped during the update step, but still stored in the lattice.

Particularly in case of regions with a high fraction of obstacle cells, it can be beneficial to divide the geometric bounding box in sub-blocks ("patches") and to allocate only those which contain fluid cells. This so-called

*Email addresses:* `markus.wittmann@rrze.uni-erlangen.de` (M. Wittmann), `thomas.zeiser@rrze.uni-erlangen.de` (T. Zeiser), `georg.hager@rrze.uni-erlangen.de` (G. Hager), `gerhard.wellein@rrze.uni-erlangen.de` (G. Wellein)

*patch-based* approach has been successfully applied in e.g. [3, 4].

If the simulation domain contains a high fraction of scattered solid matter, like in porous media, a *sparse representation* can be used where only the PDFs of fluid cells are stored [1, 9, 10, 12, 13, 15] in a one-dimensional array along with an adjacency list. This sparse "unstructured" representation requires indirect addressing, but adds a lot of flexibility for processing the nodes and allows arbitrarily shaped interfaces between MPI partitions (i.e., no need for planar boundaries). As a side effect, the bounce back boundary condition can be implicitly applied and needs no further handling.

In our work, we use the ILBDC (International Lattice Boltzmann Development Consortium) code [15], which implements a sparse representation of the lattice Boltzmann PDFs. It is designed to efficiently simulate in 3-D the flow through porous media with a low fraction of fluid space. Typical examples are packed bed reactors where a tube is randomly filled with, e.g., spheres or cylinders. Domain decomposition within the flow solver is straightforward once the sparse "unstructured" representation using the 1-D adjacency list is available and independent of the geometric structure of the simulation domain: When running with $N$ MPI processes, the 1-D adjacency list traversing all fluid cells is cut into $N$ equal-sized chunks (some may be smaller by one cell). Each process reads its chunk, determines the interface cells (i.e. cells with direct contact to other partitions), and allocates ghost cells (i.e. "duplicated cells from remote") for them. As long as the computational work per cell is constant, load imbalance does not occur. If the adjacency list was constructed with locality aspects in mind, also the amount of communication (i.e., the surface areas between adjacent partitions) is under control. Thus, the challenge is the assembly of the sparse representation and its adjacency list.

## 3. Distributed assembly of the sparse representation

The *preprocessor* assembles the adjacency list out of a full representation (fluid cells and obstacles) of the simulation domain. At startup, nothing is known about the domain except its size. Consequently it must be fully allocated to be analyzed. Through a geometric decomposition, each partition is assigned to a preprocessor rank as shown in Fig. 1 (a). Without this parallel setup, domains exceeding a single node's memory would require a large shared memory machine.

In the adjacency list each fluid cell needs a *domain-wide unique contiguous* index $I_c(x, y, z)$ with $I_c(x, y, z) \in [1, \ldots, N_f]$, where $N_f$ is the number of the fluid cells in the simulation domain. It also holds true $I_c(x_1, y_1, z_1) \neq I_c(x_2, y_2, z_2)$, when $(x_1, y_1, z_1)$ and $(x_2, y_2, z_2)$ denote two different fluid cells. Solid cells will receive $I_c(x, y, z) = 0$, because they are not considered in the list.

Several steps are required to finally obtain the domain-wide uniq contiguous index:

*Numbering scheme.* In the first step each cell receives a *domain-wide unique* but not necessarily contiguous index $I(x, y, z)$ by a numbering scheme, which is also known as index function. This scheme can be a lexicographic ordering of the cells with or without blocking, a discrete space-filling curve like the Z curve (Morton order) [8], a Hilbert curve [6], or any other positive discrete injective function $f(x, y, z)$. Hereby all cells, fluid and solid, are considered. After numbering, the fluid cell index contains gaps, which correspond to the indices of the solid cells. During this step the indices $I$ and the fluid cell coordinates are stored in the adjacency list.

*In- and outcells.* In the next step gaps in the index $I$ are removed. Each rank determines its *incells* $\mathcal{I}$ and *outcells* $\mathcal{O}$. Incells are cells with domain-wide unique index denoted by $I(\mathcal{I}_n)$ whose previous cell with index $I(\mathcal{I}_n) - 1$ is located on another partition. Each incell $\mathcal{I}_n$ corresponds to an outcell $\mathcal{O}_n$. The latter is the first cell with $I(\mathcal{O}_n) > I(\mathcal{I}_n)$ that does not reside on the same partition as $\mathcal{I}_n$. Fig. 1(a) shows two incells and outcells of rank 1.

*Contiguous index generation.* Each process maintains an array of a data structure called *incell info* or short `ici`. Each `ici` stores for each incell $\mathcal{I}_n$ its index $I(\mathcal{I}_n)$, the index $I(\mathcal{O}_n)$ of the corresponding outcell, the number of fluid cells $N_f(\mathcal{I}_n)$ with index $I(x, y, z)$ for which $I(\mathcal{I}_n) \leq I(x, y, z) < I(\mathcal{O}_n)$, the MPI rank of the local process, and an entry which will later receive the globally contiguous index $I_c$ of the first fluid cell in the range of this `ici`.

The partitions are considered as the leaves of an (incomplete) octree. From bottom to top on each level $l$ of the tree a "sibling master" is selected among each group of up to eight children. This master receives its siblings' `ici` arrays and stores them in a local array $A_l$. In a new array $B_l$ a copy of $A_l$ is stored and sorted by the domain-wide unique index $I(\mathcal{I}_n)$ of the `ici`s. For all `ici`s the rank is set to the sibling master's one. Two `ici`s in array $B_l$ are merged if the outcell index $I(O_i)$ of `ici` $B_l(i)$ equals the incell index $I(\mathcal{I}_{i+1})$ of the following `ici` $B_l(i+1)$. These two elements are replaced by a new one with the incell index $I(\mathcal{I}_i)$ of $B_l(i)$ and the outcell index $I(\mathcal{O}_{i+1})$ of $B_l(i+1)$. The number of fluid cells associated with $B_l(i)$ and $B_l(i+1)$ are summed up. This procedure is repeated until the top level is reached, i.e., when no siblings are left.

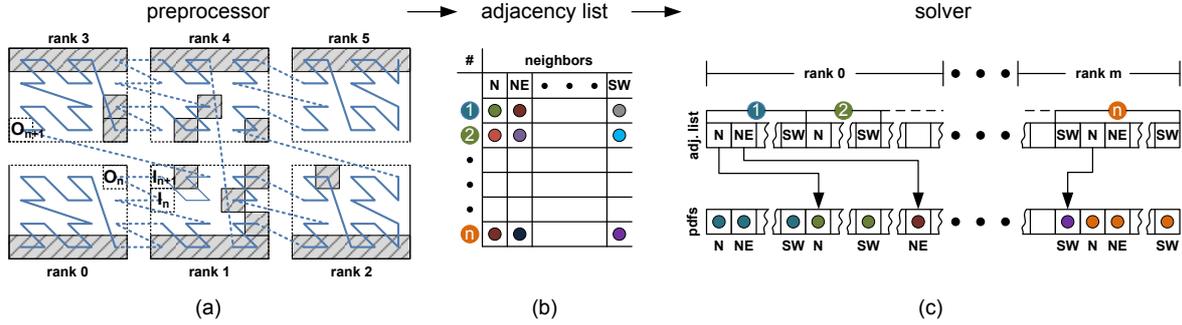

Figure 1: The simulation domain is geometrically decomposed by the preprocessor and partitions are assigned to processes (a). The preprocessor sets up the adjacency list with a domain-wide unique contiguous index (b). This list is picked up by the flow solver and cut into equally sized chunks so that each process has the same number of fluid cells (c).

The top level process computes the domain-wide unique contiguous indices $I_c$ for all icis in its $B_l$ array. The index $I_c$ starts at 1 for the first entry $B_l(1)$ and is incremented for each further entry by the number of fluid cells $N_f(\mathcal{I}_{n-1})$ of the previous entry $B_l(n-1)$. These new indices can be mapped back for each ici to the icis in $A_l$. Array $A_l$ is sorted by the rank, and subsequently contiguous chunks of entries are sent back to their origin ranks.

The receiving rank overwrites its $B_l$ array with the received data and repeats the steps of mapping back indices and sending back icis until all leaves have been reached.

Each process can now assign a domain-wide unique contiguous index $I_c$ to the fluid cells in the adjacency list and in the full representation. The missing adjacency information can be obtained by inspecting the neighborhood of each fluid cell in the full representation. Since the neighborhood may contain cells outside the local partition, halo layers are needed.

*Storing the adjacency information.* Prior to writing the adjacency list (Fig. 1(b)) to disc it is sorted in parallel according to the unique contiguous index $I_c$. Each process then writes an equally sized chunk of the list. This results in a better IO performance than letting each process write scattered parts. The information stored in the adjacency file is later used by the solver (Fig. 1(c)).

## 4. Preprocessing with graph partitioning

The flow solver picks up the adjacency list for simulation and cuts it into equally sized chunks as shown in Fig. 1(c). No respect is taken to the shape of the resulting partitions. Graph partitioning algorithms could generate more well-formed partitions. Therefore the widely used graph partitioner METIS [7] and PT-SCOTCH [2] were employed for comparison.

The full representation of the simulation domain is transformed into a graph. Each fluid node is considered as a vertex. PDFs represent edges between the fluid cells they origin from and the fluid node they are pointing to. Utilizing all PDFs, i.e. the complete neighborhood of a fluid cell, is called the "full neighborhood" (FN) or Moore neighborhood. It is possible to reduce the number of edges of the graph by only using PDFs along the main coordinate axes, which we name "reduced neighborhood" (RN) or von Neumann neighborhood.

To avoid confusion with the lattice partition that is owned by a preprocessor, we use the term "graph partition" for the partition assignment generated by a graph partitioner.

*METIS.* For METIS (version 4.0.3, k-way, FN) the graph is built serially by the preprocessor and given to METIS for partitioning. Subsequently the preprocessor determines the number of fluid cells belonging to each graph partition and computes the domain-wide unique contiguous start index $I_c$ of each graph partition.

It is iterated in a lexicographic order (with blocking) over the cells of the full representation. The graph partition a fluid cell belongs to is determined and the next free index $I_c$ inside this graph partition is assigned. After all cells have been processed separate files containing the sparse representation for each graph partition are written to disc and picked up later by the flow solver.

*PT-SCOTCH.* With PT-SCOTCH (version 5.1.12, RN and FN) partitioning is performed in parallel. The graph has to be setup in a parallel and distributed manner by the preprocessor. After partitioning, as with METIS, the number of cells of each graph partition has to be determined and domain-wide unique contiguous start indices must be computed.

A global lexicographic ordering (with blocking) over all cells in the simulation domain is used to visit each cell in the local partition. Index assignment is thereby performed in the same way as with METIS. Graph partitions can spread across more than one process. In this case each process has to determine how many cells of which graph partition are located on its subdomain. The domain-wide unique contiguous index $I_c$ is assigned to fluid cells of such graph partitions in contiguous blocks for each process.

The $N$ processes of the flow solver would later pick up the generated sparse representation and cut it into $N$ equally sized chunks. But the generated graph partitions from PT-SCOTCH might not all have the same number of cells and therefore would not correspond with the partitions gained by simple cutting. The sparse representation therefore is extended by a list of the domain-wide unique contiguous start indices of each graph partition for recovering them at the initialization of the solver.

*Note:* The goal here is not to compare METIS and PT-SCOTCH. Results for these graph partitioners should only give an estimate of the performance that can be achieved with these tools, since graph partitioning has the disadvantage that the processing order of the cells is not defined, which can have enormous impact on performance.

Further the behavior of PT-SCOTCH can be influenced by complex *strategy strings* specifying when and which partitioning method should be applied. Here just the default strategy string was used, which according to [2] should yield good results in most cases. Also notable is that PT-SCOTCH produces different partitions when different numbers of processes for the same geometry and the same number of partitions are used.

For the usage of graph partitioning the number of flow solver processes used for CFD simulation later must already be known at preprocessing time, as for this number of processes partitions are generated. This restriction does not apply for our method.

## 5. Test bed

An overview of the two systems used as test bed is given in Tab. 1. The first system "Westmere" is a cluster with fully nonblocking QDR InfiniBand (IB) interconnect. The four-socket "MagnyCours" system is a large shared memory machine.

|                | **Westmere** | **MagnyCours** |
|----------------|---|---|
| Processor      | Intel Xeon X5650 | AMD Opteron 6176 |
| Freq. [GHz]    | 2.67 | 2.30 |
| max. Turbo [GHz] | 3.06 | – |
| Cores          | 6 | 2 x 6 |
| L1 [kB]        | 32 | 64 |
| L2 [kB]        | 256 | 512 |
| L3 [MB]        | 12 | 5 |
| Sockets        | 2 | 4 |
| NUMA domains   | 2 | 8 |

Table 1: Details of the two systems Westmere and MangyCours used for preprocessing geometries and performance evaluation of the flow solver. The Westmere system is a cluster with QDR InfiniBand as interconnect.

## 6. Resource requirements for preprocessing

The preparation of the simulation domain ("preprocessor") and the flow simulation are separated in ILBDC. In the first step the simulation is prepared, i.e. the sparse representation is extracted out of the full representation of the domain. In the second step the flow simulation takes place.

This section examines the resource requirements like run time and memory usage of the preprocessing. The next section then evaluates the performance of the CFD simulation itself.

An empty channel (called channel) and a tube filled with a sphere packing (called packing) as in Fig. 2, were chosen as benchmark geometries. The empty channel exhibits nearly only fluid cells in contrast to the packing with a high fraction of solid cells and a complex structure. Both geometries can be scaled to any diameter $d$ with dimensions of $(5*d) \times d \times d$ cells.

We use geometries with diameters $d$ = 500, 800, and 1000. The duration of the preprocessing depends on the dimensions and the ratio of fluid cells of the simulation domain. The "larger" a geometry becomes the more observable the runtime becomes, which ranges from minutes to hours.

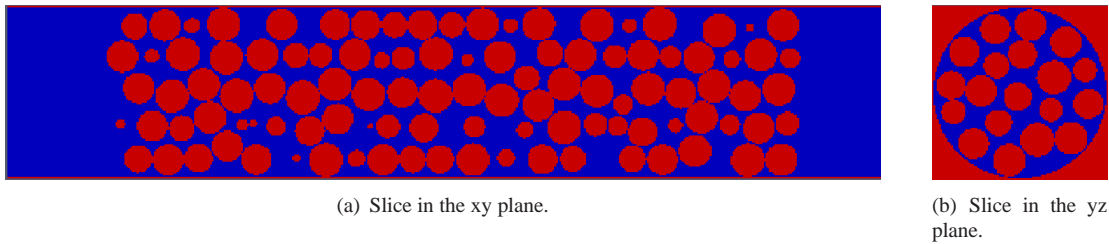

(a) Slice in the xy plane.

(b) Slice in the yz plane.

Figure 2: Slices of a packed bed reactor of spheres. Blue indicates fluid cells and red solid cells.

| method | nodes | PPN | memory [GB] | | duration [s] | | IO [GB] |
|---|---|---|---|---|---|---|---|
| | | | max/proc. | total | w/o IO | w/ IO | |
| *channel, d = 800, ≈ 2, 5 ∗ 10⁹ fluid cells* | | | | | | | |
| lex. ordering 1 | 128 | 1 | 7.1 | 908 | 84 | 883 | 371 |
| lex. ordering 100 | 128 | 1 | 7.1 | 907 | 53 | 834 | 371 |
| PT-SCOTCH | 128 | 1 | 10.8 | 1,298 | 800 | 3901 | 371 |
| lex. ordering 1 | 256 | 12 | 2.4 | 1,469 | 57 | 965 | 371 |
| lex. ordering 100 | 256 | 12 | 2.4 | 1,003 | 14 | 1479 | 371 |
| *packing, d = 800, ≈ 1, 1 ∗ 10⁹ fluid cells* | | | | | | | |
| lex. ordering 1 | 128 | 1 | 4.1 | 413 | 65 | 382 | 163 |
| lex. ordering 100 | 128 | 1 | 4.1 | 411 | 38 | 345 | 163 |
| PT-SCOTCH | 128 | 1 | 8.0 | 638 | 542 | 2092 | 163 |
| lex. ordering 1 | 256 | 12 | 2.1 | 1,130 | 66 | 475 | 163 |
| lex. ordering 100 | 256 | 12 | 2.1 | 506 | 22 | 433 | 163 |

Table 2: Resource requirements for preprocessing an empty channel ($\approx 2, 5 * 10^9$ fluid cells) and a packed bed of spheres ($\approx 1, 1 * 10^9$ fluid cells) with dimensions of $4000 \times 800 \times 800$ cells each on 128 (1 PPN) and 256 (12 PPN) compute nodes of the Westmere cluster with lexicographic ordering and PT-SCOTCH. With PT-SCOTCH preprocessing was neither possible on 128 nodes with 12 PPN nor on 256 nodes with 1/12 PPN. Runtime is dominated by disc IO to the parallel file system (Lustre), which achieved between 100 and 550 MB/s depending on the load on the file system.

As an example we selected channel and packing with dimensions of $4000 \times 800 \times 800$ cells each, which are $\approx 2, 5 * 10^9$ and $\approx 1, 1 * 10^9$ fluid cells, respectively.

Table 2 shows details for preprocessing these two geometries with our method (lexicographic ordering with blocking factor 1 and 100) and with PT-SCOTCH. Obviously the runtime is dominated by IO, i.e. writing the sparse representation to disc. The time for index generation (w/o IO) here becomes negligible. Lexicographic ordering and PT-SCOTCH show heavily different durations. Despite the difference in non-IO durations, the main cause is the varying IO bandwidth. The underlying parallel file system (Lustre) on the Westmere cluster is designed for an aggregated IO bandwidth of 3 GB/s but the bandwidth a particular job actually can see heavily depends on the load generated by other users and their applications. Typically, we only could sustain 100 – 500 MB/s with our application in non-dedicated cluster operation.

Comparing the memory usage when using 128 nodes and 1 process per node (PPN) for preprocessing reveals differences between lexicographic ordering and the usage of the graph partitioner. Fig. 3 shows a recorded memory usage trace (total memory usage over all used nodes) during preprocessing the packing geometry ($d = 800$) with lexicographic ordering (blocking factor 1 and 100) and PT-SCOTCH.

PT-SCOTCH was still partitioning after 12 hours with 128 nodes and 12 PPN nor with 256 nodes and 1 or 12 PPN. Comparing lexicographic ordering for 128 and 256 nodes shows that with a reasonable blocking factor like 100 the non-IO time decreases with more nodes. With more processes the total amount of memory used increases, because each process requires its own work-ing set. Also with blocking factor 1 larger interfaces are generated than with blocking factor 100. Therefore larger data structures are required to describe this information which becomes significant as the number of processes increases. This explains the huge difference of total memory usage for lexicographic ordering with blocking factor 1 and 100 in Tab. 2.

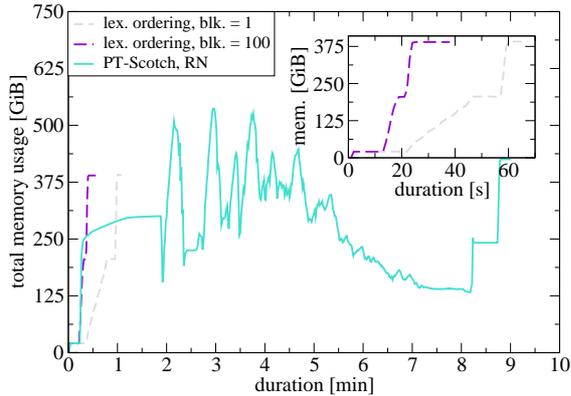

Figure 3: Memory usage traces for preprocessing a packed bed of spheres with dimensions $4000 \times 800 \times 800$ cells $\approx 1, 1 * 10^9$ fluid cells with lexicographic ordering and PT-SCOTCH.

For PT-SCOTCH it seems favorable to use as few processes as possible for partitioning. With, e.g. 256 nodes and 1 or 12 PPN (256 or 3072 processes), partitioning geometries with diameters $\geq 500$ did not finish after 12 hours. As an alternative we used 128 nodes to have enough total memory available but use only 1 PPN to reduce the number of processes. The reduced neighborhood was used to reduce the required memory for representing the graph.

## 7. Performance of CFD simulation

The benchmarked implementation of the flow solver uses double precision floating-point for representing the PDFs, a *structure-of-arrays* data layout, *pull* scheme [14], and *nontemporal* stores for writing the new PDFs to the destination grid [15]. One solver process was bound to each physical core, i.e., twelve MPI-processes per node on Westmere and 48 on MagnyCours were used. As performance metric we use *fluid lattice updates per second* (FLUP/s) which only takes the updated fluid cells into account. One lattice update involves approx. 200 floating point operations (FLOPs) in our implementation. This number was extracted by inspecting the generated assembler source of the LB kernel.

For evaluating the single node performance and strong scaling we used the channel and packing geometries with dimensions of $500 \times 100 \times 100$ cells, which contain $\approx 4, 8 * 10^6$ and $\approx 2, 1 * 10^6$ fluid cells, respectively.

*Single node performance.* The performance of ILBDC depends heavily on the chosen numbering scheme used for setting up the adjacency list. Figure 4 shows that the performance varies by about 20% depending on the blocking factor (black curves). A blocking factor of one is equivalent to no blocking at all.

Comparing data in Fig. 4(a) and 4(b) reveals that the geometry has no significant impact on performance, although the adjacency of cells and the number of fluid cells is different. The implementation of the discrete space-filling Z curve uses bitwise interleaving of cell coordinates (1 bit) or interleaves two-bit groups of the coordinates (2 bit), which in general results in better performance. METIS and PT-SCOTCH give a comparable performance which is always in the range of the maximum performance achieved with lexicographic ordering.

*Strong scaling.* Using more than eight nodes of the Westmere cluster the two benchmark geometries show nearly the same performance characteristics, albeit on different levels (Fig. 5(a)). This may be explained by the different number of fluid cells. It also reveals that choosing the right blocking factor for lexicographic ordering has more influence on performance than on a single node. The data in Fig. 5 shows a 50% performance fluctuation on 64 nodes depending on the blocking factor. With 128 nodes no more performance is gained since the number of cells each process owns is as low as $\approx 1400$ fluid cells. Utilizing even more nodes is counterproductive.

Lexicographic ordering with blocking outperforms the Z curve (1 and 2 bit) as well as partitioning with METIS as shown in Fig. 5(b). With METIS an early performance decrease of one third compared to other methods is observable at already 128 nodes. PT-SCOTCH generates excellent results for this geometry, in some cases even a slightly better performance than lexicographic ordering with blocking.

*Large geometries.* For the evaluation of the performance of "larger" geometries we used 256 nodes of the Westmere cluster with 12 PPN resulting in 3072 processes. Figure 6 shows the performance achieved with lexicographic ordering (blocking factor 1 and 100) and partitioning with PT-SCOTCH for the benchmark geometries channel and packing with diameters $d =$

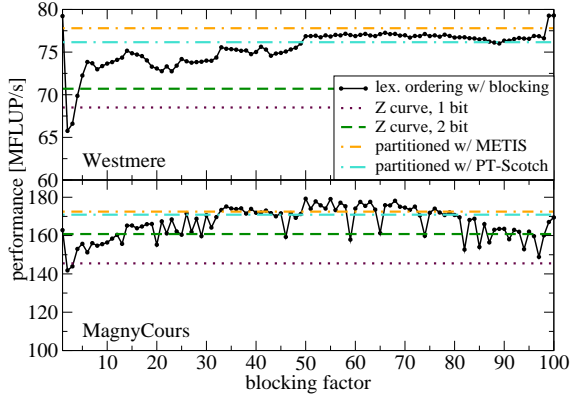

(a) Empty channel: $500 \times 100 \times 100$.

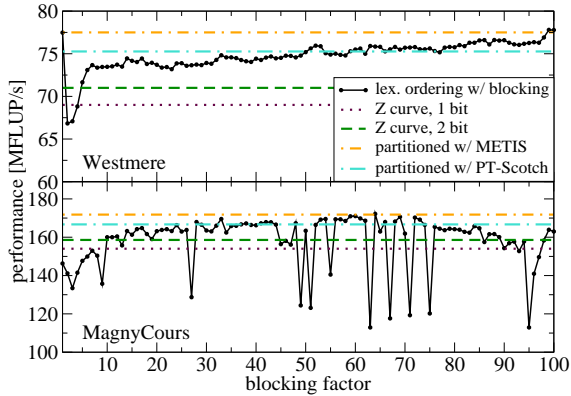

(b) Tube w/ packed bed of spheres: $500 \times 100 \times 100$.

Figure 4: Pull scheme with different numbering schemes: lexicographic ordering with blocking (black curves), Z curve (1 and 2 bit, purple and green curves), and partitioned with METIS (orange curves) on one node of Westmere and MagnyCours (twelve and 48 MPI processes, respectively).

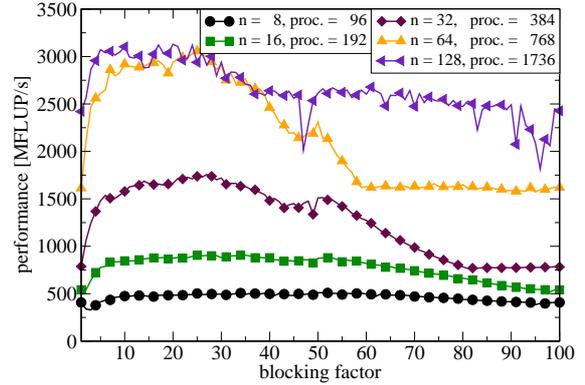

(a)

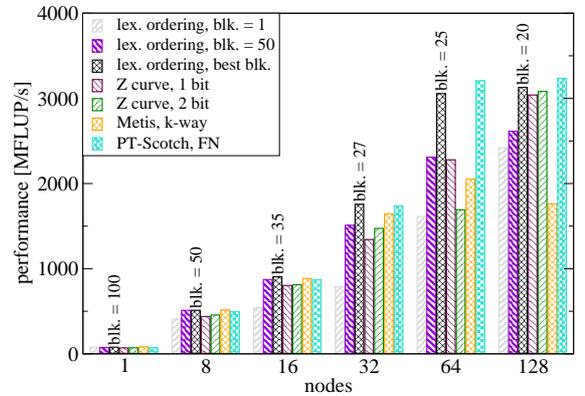

(b)

Figure 5: Strong scaling on 8 – 256 Westmere nodes (12 PPN) for a packed bed of spheres (packing) with dimensions of $500 \times 100 \times 100$ cells $\approx 2, 1 * 10^6$ cells for lexicographic ordering with blocking (a) and compared to Z curve 1 and 2 bit, and METIS (b). Dramatic differences in performance make the right choice of the blocking factor for adjacency list setup crucial.

500, 800, and 1000 cells. The dimensions of both geometries are $(d * 5) \times d \times d$ cells.

Lexicographic ordering (blocking factor 100) and PT-SCOTCH achieve nearly the same results except for packing $d = 800$. Here the performance of PT-SCOTCH decreases significantly. The cause might be the complex geometry in combination with the reduced neighborhood which was used for setting up the graph.

The importance of the right blocking factor for lexicographic ordering can be seen for both geometries. Over 200 % performance is gained by switching from blocking factor 1 to 100. This can be explained by analyzing the generated partitions.

Therefore we have analyzed the number of neighbor partitions and the number of remote links per partition. The latter one is the number of PDFs of adjacent partitions that must be communicated.

With a blocking factor of 1 the domain is cut into (partial) slices. This causes a low count of neighbor partitions (Fig. 7(a) left panel), but also increases inter-partition communication (Fig. 7(b) left panel). For a larger blocking factor, e.g. 100, the number of neighbor partitions is increased (Fig. 7(a) middle panel), but the number of links to communicate decreases (Fig. 7(b) middle panel).

Comparing the partitions gained by lexicographic ordering with a blocking factor of 100 and PT-SCOTCH (Fig. 7(a) and 7(b) middle and right panels) more partitions with a significantly increased number neighbor partitions (50 – 150) were generated.

PT-SCOTCH generated for a packed bed of spheres with dimensions $5000 \times 1000 \times 1000$ cells a graph par-tition (Fig. 8) that had a lot of unconnected cells spread over nearly the complete simulation domain. Due to a restriction in the flow solver we are not able to use partitions for which the product of the boudning box dimensions exceeds $2^{31}$ cells.

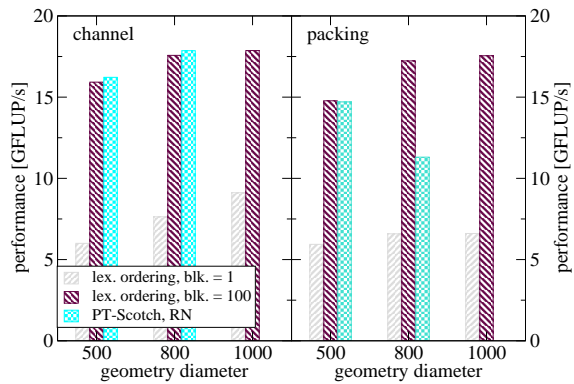

Figure 6: Performance of the flow solver on 256 Westmere nodes (12 PPN) for the channel and packing benchmark geometries preprocessed with lexicographic ordering (blocking factor 1 and 100) and partitioning with PT-SCOTCH. The geometries' dimensions are $(5 * d) \times d \times d$ cells, where $d$ denotes diameter of the geometry. For channel $d = 1000$ we were not able to generate a graph partitioning with PT-SCOTCH.

## 8. Conclusion and future work

We presented an algorithm to preprocess a geometry for flow simulation that can compete with established graph partitioning tools. It provides the advantage that the number of flow solver processes does not need not to be known at preprocessing time. This number can also be freely chosen at every restart of the simulation. Independently of the number of processes used for preprocessing a specific geometry and a chosen numbering scheme always produce exactly the same sparse representation.

Choosing the right blocking factor for lexicographic ordering is crucial for good performance of simulations based on the sparse representation of the simulation domain. Unfortunately the best choice depends on the machine architecture, the number of fluid nodes, the adjacency of the nodes, the number of processes, and the communication network.

We are working on a heuristic approach that arrives at reasonable predictions for the best blocking factor. Also we are investigating an auto-tuning approach which determines the right parameter set for a given geometry and architecture by only simulating samples of the geometry.

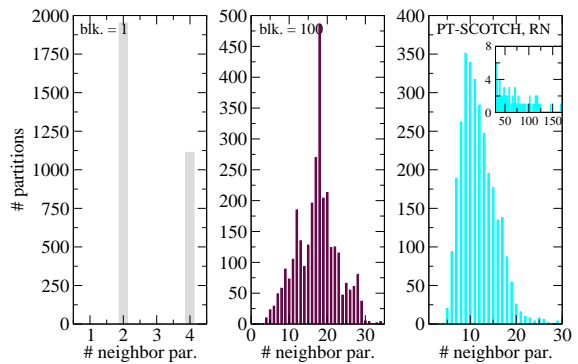

(a) Number of neighbor partitions. Axis labels of the inset are the same as of the surrounding graph one's.

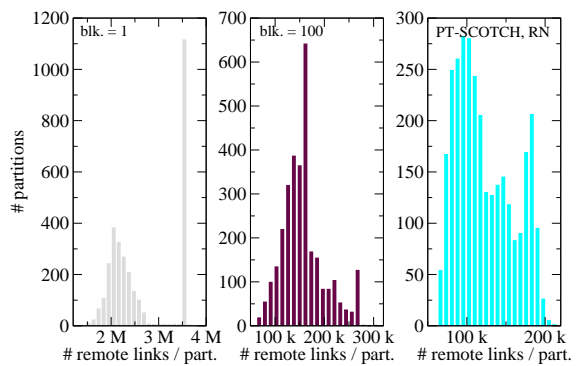

(b) Number of remote links per partition.

Figure 7: Histograms over the number of neighbor partitions (a) and over the number of remote links per partition (b), i.e. the number of adjacent fluid cells' PDFs belonging to a different partition.

*Acknowledgements.* Lively discussions with Jan Treibig, and Johannes Habich are gratefully acknowledged. This work was supported by BMBF under grant No. 01IH08003A (project SKALB).

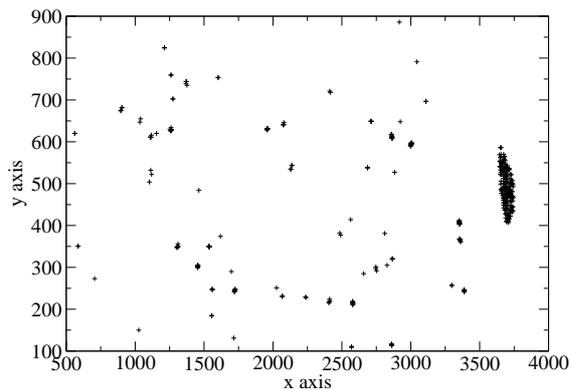

Figure 8: Top view of graph partition consisting of a lot of unconnected cells spread over the simulation domain. It was generated by creating 3072 partitions with PT-SCOTCH out of the packing geometry (5000 × 1000 × 1000 cells).